\documentclass[useAMS,usenatbib]{mn2e}
\usepackage{natbib}
\usepackage{epsf}
\usepackage{graphicx}
\usepackage{textcomp}

\def\simlt{\mathrel{\rlap{\lower 3pt\hbox{$\sim$}}
        \raise 2.0pt\hbox{$<$}}}
\def\simgt{\mathrel{\rlap{\lower 3pt\hbox{$\sim$}}
        \raise 2.0pt\hbox{$>$}}}

\title[CMDs to separate OIR disc/jet emission]{A tool to separate optical/infrared disc and jet emission in X-ray transient outbursts: the colour-magnitude diagrams of XTE J1550--564
}
\author[D. M. Russell et al.]{D. M. Russell$^{1}$\thanks{E-mail: d.m.russell@uva.nl}, D. Maitra$^{2}$, R. J. H. Dunn$^{3}$, R. P. Fender$^{4}$ \\
$^{1}$Astronomical Institute `Anton Pannekoek', University of Amsterdam, P.O. Box 94249, 1090 GE Amsterdam, the Netherlands\\
$^{2}$Department of Astronomy, University of Michigan, 500 Church Street, Ann Arbor, MI 48109, USA\\
$^{3}$Technische Universit\"at M\"unchen, Excellence Cluster Universe, Boltzmannstrasse 2, D-85748 Garching, Germany\\
$^{4}$School of Physics \& Astronomy, University of Southampton, Highfield, Southampton, SO17 1BJ, UK\\
}
\begin{document}


\pagerange{\pageref{firstpage}--\pageref{lastpage}} \pubyear{2010}

\maketitle

\label{firstpage}

\begin{abstract}
It is now established that thermal disc emission and non-thermal jet emission can both play a role at optical/infrared (OIR) wavelengths in X-ray transients. The spectra of the jet and disc components differ, as do their dependence on mass accretion properties. Here we demonstrate that the OIR colour-magnitude diagrams (CMDs) of the evolution of the X-ray transient XTE J1550--564 in outburst can be used to separate the disc from the jet. Monitoring in two wavebands is all that is required. This outburst in 2000 was well studied, and both disc and jet were known to contribute. During the outburst the data follow a well defined path in the CMD, describing what would be expected from a heated single-temperature blackbody of approximately constant area, except when the data appear redder than this track. This is due to the non-thermal jet component which dominates the OIR moreso during hard X-ray states at high luminosities, and which is quenched in the soft state. The CMD therefore shows state-dependent hysteresis, in analogy with (but not identical to) the well established X-ray hardness--intensity diagram of black hole transients. The blackbody originates in the X-ray illuminated, likely unwarped, outer accretion disc. We show that the CMD can be approximately reproduced by a model that assumes various correlations between X-ray, OIR disc and OIR jet fluxes. We find evidence for the OIR jet emission to be decoupled from the disc near the peak of the hard state.
\end{abstract}


\section{Introduction}

Historically, OIR emission from low mass X-ray binaries (LMXBs) during outburst is known to originate in the outer regions of the accretion disc, mostly due to the reprocessing of UV/X-ray photons \citep*{franet02,charco06}. Non-thermal emission, or thermal cyclosynchrotron in the OIR regime has also been reported since the 1980s \citep{fabiet82,motcet83} but only recently has its origin been identified as most likely synchrotron emission from a compact jet, an outflow and not an inflow, due to its spectral energy distribution \citep[SED; e.g.][]{corbfe02,miglet06,hyneet06} high brightness temperature and rapid variability which is correlated with X-ray variability in a complex manner \citep*[e.g.][]{kanbet01,malzet04,caseet10}, long-term (anti-) correlations with X-ray flux \citep[e.g.][]{russet06,coriet09} and linear polarization (\citealt{shahet08,russfe08}; for a review see \citealt{russfe10}).

If one wishes to study the disc, or the jet, at OIR wavelengths it is usually necessary to isolate the desired component, since contamination from the other can affect the observed spectrum, variability properties (e.g. fractional rms) and correlations with other wavelengths. If the jet is to make a contribution in quiescence \citep[which it may do in some cases; see][]{cantet10,maitet10} this will affect estimates of black hole or neutron star mass, since these rely on the successful isolation of the companion star emission at these low luminosities \citep{casa07}. It is therefore of utmost importance to isolate the disc, jet and star emission, which can be achieved by studying and modelling their different spectra \citep*[e.g.][]{market05}, their variability \citep[e.g.][]{gandet10} or their long-term evolution \citep{russet10}.

In \cite{maitba08}, it was illustrated that OIR colour-magnitude diagrams (CMDs) of LMXB outbursts (the OIR equivalent of X-ray hardness intensity diagrams; HIDs) have the potential to separate thermal and non-thermal emission. For this, monitoring of the source in just one optical and one near-infrared (NIR) band was required. The authors were able to reproduce the correlation between OIR colour and magnitude in eight outbursts of the neutron star LMXB Aql X--1 using a model consisting of a simple single-temperature accretion disc heated up. This simple but elegant technique identified the dominating process as X-ray reprocessing on a disc of constant area, and ruled out emission from the jet and the viscously heated disc \citep{maitba08}. Synchrotron emission from the jet is expected to have an OIR spectral index that does not depend on luminosity \citep[$\alpha \approx -0.7$ for optically thin synchrotron, where $F_{\nu} \propto \nu^{\alpha}$; $\alpha \approx 0.0$ for optically thick, self-absorbed synchrotron;][]{blanko79,fend01}. For X-ray reprocessing, $-1 \leq \alpha \leq +2$ is expected in a given waveband ($\alpha = +2$ is the Rayleigh-Jeans tail of the blackbody); the equivalent for a viscous disc is $+\frac{1}{3} \leq \alpha \leq +2$ \citep{franet02}.

The 2000 outburst of XTE J1550--564 is the only one for which optical, NIR and X-ray data were obtained throughout the entire outburst with state transitions. In the X-ray regime, the source made a transition from a hard state to a soft/intermediate state \citep[in the context of jet activity the source was in a soft state; see discussion in][]{russet07b} and back, at lower luminosity, to a hard state \citep*[e.g.][]{rodret03}, obeying the well known hysteresis of black hole transients in the HID \citep*[e.g.][]{maccco03,fendet04,fendet09,bell10,dunnet10}. It is clear that there is a quenching and recovering of the NIR flux (and the optical to a lesser extent) contemporaneously with the X-ray softening and hardening, respectively \citep[see][]{jainet01}. Synchrotron emission from the jet produces the NIR flux during the hard state; this jet is quenched during the soft state \citep[e.g.][]{jainet01,corbet01}. \cite{russet07b} showed that the NIR emission from the jet is brighter during the hard state decline than during the hard state rise, at a given X-ray luminosity (NIR--X-ray hysteresis), which may be due to a more powerful jet on the decline \citep[which also likely dominated the X-ray flux;][]{russet10}, or changes in the jet efficiency and/or the disc viscosity.

In this paper we use the OIR data of XTE J1550--564 during its 2000 outburst \citep{jainet01} to construct its CMD. V and H-bands are used; the spectral indices are measured between 550 and 1660 nm (a range of 0.5 dex in frequency or wavelength). We also construct an additional CMD using I and H-bands. We demonstrate that these diagrams can be used as diagnostic tools to constrain the disc and jet contributions. We infer some properties of both the jet and disc emission of XTE J1550--564 and the relationship between them using theoretical scaling models. Preliminary results, including other sources (black holes and neutron stars) were presented in \cite{russet08}, and in a paper in preparation we will analyze the CMDs of a large collection of sources and define global properties and trends.  Here we demonstrate the technique using the CMDs of XTE J1550--564 during its 2000 outburst, which is the only BHXB outburst in the literature with state transitions and with dense optical and NIR monitoring.

In section 2 we present the CMD and the blackbody model. In section 3 we analyze and discuss the results in the context of the dominating OIR emission mechanisms and the relation between the CMD and HID. We attempt to reproduce the observed evolution through the CMD using the expected theoretical relations between the jet and irradiated disc components in section 4. We summarize our conclusions in section 5.

\begin{figure*}
\centering
\includegraphics[width=15cm,angle=270]{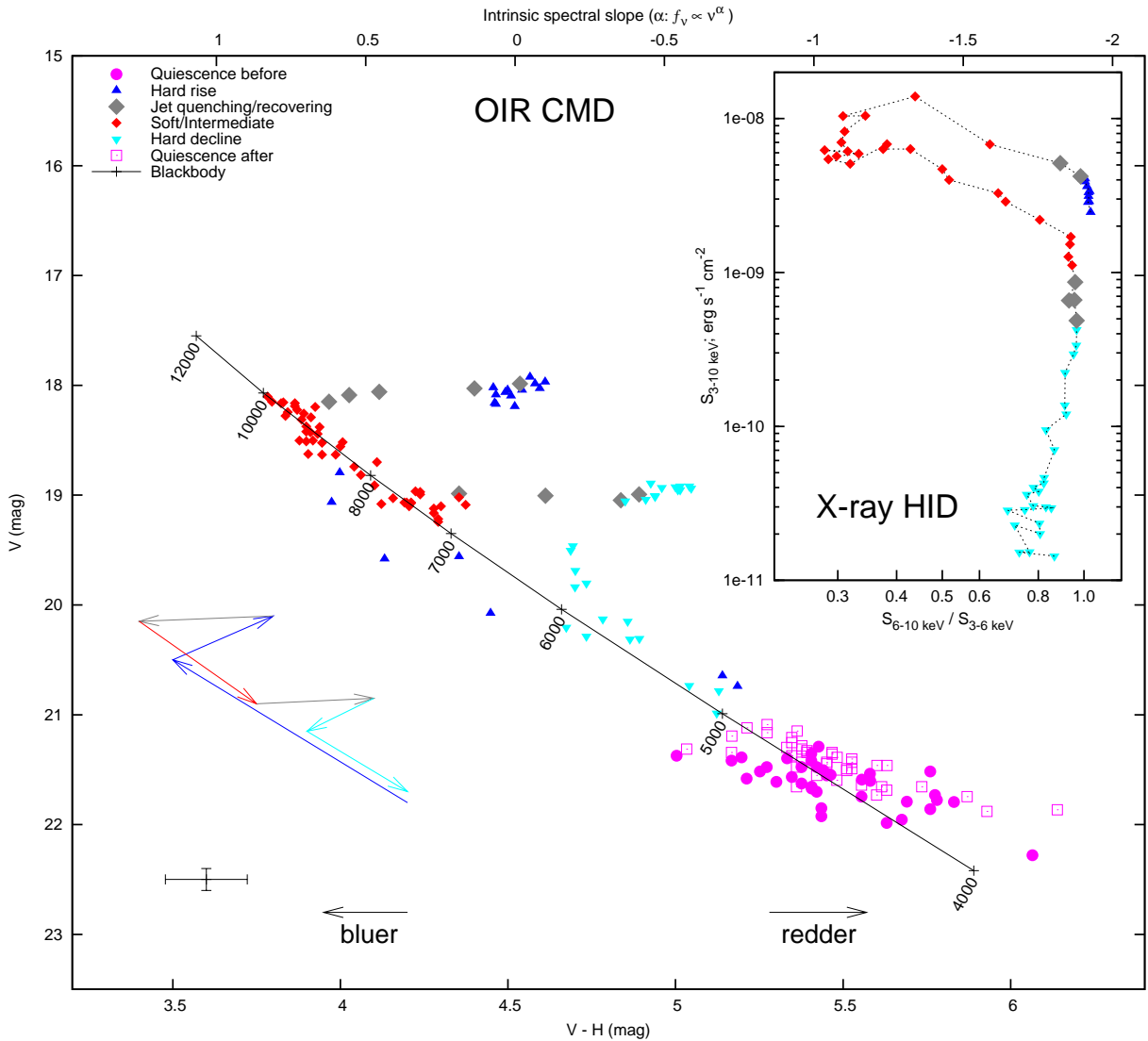}
\caption{OIR CMD of the 2000 outburst of XTE J1550--564 using V and H-bands, with X-ray HID as an inset. See text for explanations and analysis.}
\end{figure*}

\section{The colour-magnitude diagram}

The CMD using V and H-bands is presented in Fig. 1. Error bars are shown in the lower left of the CMD, which represent the mean magnitude errors quoted in the original data sources, and colour errors as propagated from these values. The intrinsic spectral index $\alpha_{\rm int}$ is estimated from the known interstellar extinction $A_{\rm V}$ towards the source and the extinction law of \cite*{cardet89}.

The extinction towards XTE J1550--564 was first estimated to be low; $A_{\rm V} = 2.2 \pm 0.3$ based on the equivalent width of interstellar absorption lines in the optical spectrum \citep{sancet99}. However, based on the neutral hydrogen column density estimated via X-ray spectral fitting \citep*{tomset01,tomset03,kaaret03}, the likely value is $A_{\rm V} \sim 5.0$. Recently, \cite{oroset11} derived extinction values in the NIR of $A_{\rm J} = 1.17 \pm 0.11$, $A_{\rm H} = 0.762 \pm 0.074$ and $A_{\rm K} = 0.507 \pm 0.050$, which are consistent with $A_{\rm V} \sim$ 4--5 \citep{cardet89}. Here we adopt $A_{\rm V} = 5.0$, and this results in a blue ($\alpha > 0$) OIR SED during outburst; the SED is too red if the previously estimated $A_{\rm V} \simlt 2.5$ is used \citep[see discussion in][]{russet07b}. Following equation 1 of \cite{maitba08}, $A_{\rm V} = 5.0$ gives an intrinsic spectral index $\alpha _{VH;\rm int} = 5.073 - 1.122 \times (V-H)_{\rm obs}$, which is shown on the upper abscissa of the CMD. X-ray data from \cite{dunnet10} are used to construct the HID, which is shown as in inset in Fig. 1.

The symbols in the CMDs depend on X-ray state and jet activity. Pink circles are data taken during quiescence before the outburst, blue upward triangles are from the hard state rise of the outburst \citep[starting from the first OIR data clearly above the quiescent level;][]{jainet01}, red diamonds are from the soft/intermediate state, cyan downward triangles are from hard state outburst decline and pink open squares are from quiescence after the outburst. State definitions are those defined by \cite{rodret03}. The OIR colour changes during state transitions indicative of the jet quenching or recovering are presented as grey diamonds. The HID is colour-coded in the same manner as the CMD. While the OIR colour change due to the jet quenching is during the hard to soft/intermediate state transition, the OIR colour change due to the jet recovering is when the source is fully back in the hard state and declining in luminosity \citep[see also][]{russet10}. For this reason we colour-code the OIR data before the jet recovery as the soft/intermediate state in Fig. 1.

\subsection{The blackbody model}

We adopt the model of \cite{maitba08} for the relationship between colour and magnitude for a single-temperature heated blackbody. This approximates what may be expected from the X-ray irradiated outer accretion disc. The intrinsic, viscously heated disc can be approximated by a multi-temperature blackbody producing a power law spectrum of spectral index $\alpha = +1/3$ \citep{franet02}. At longer wavelengths, a Rayleigh-Jeans tail of the blackbody has $\alpha = +2$ and between these two regimes we expect $+1/3 \leq \alpha \leq +2$. Likewise, the irradiated disc may result in a power law spectrum with $\alpha = -1$ \citep{kingri98,franet02} at short wavelengths, which peaks ($\alpha = 0$) then drops as a Rayleigh-Jeans tail ($\alpha = +2$) at longer wavelengths \citep[the peak has been identified in some sources; e.g.][]{hyne05}. We therefore expect $-1 \leq \alpha \leq +2$ for the irradiated disc. The model we adopt here, a single-temperature blackbody, could approximate the viscous outer disc or the irradiated outer disc, so long as the observations are not in the short wavelength power law regimes ($\alpha = +1/3$ for the viscous disc and $\alpha = -1$ for the irradiated disc).

The temperature of the blackbody in the model depends on the intrinsic colour, so temperature estimates also suffer from uncertainties that propagate from inaccurate extinction. The normalization of the model represents the apparent size of the blackbody, which we approximate to a disc of outer radius equal to the Roche lobe radius viewed at an inclination and at a given distance. The disc radius is estimated from the known orbital period and the masses of the star and the black hole from the literature. The system masses, orbital period, inclination angle and distance to the source are taken from \cite{oroset02}. We use a normalization parameter, $f$ to approximately overlap the model with the data in the CMD. We do not fit the blackbody model to the data; the values of $f$ are a function of both the uncertainties in the system parameters and/or uncertainties in the disc geometry, such as a warped disc or a disc with an outer radius different to that of the Roche lobe radius.

\section{Results and analysis}

The OIR CMD using V and H-bands (Fig. 1) shows clear tracks that progress throughout the outburst. The source is reddest (lowest spectral index) at the lowest flux levels (during quiescence). During the initial hard state rise the source became bluer (the spectral index increased) and followed a path approximating a heated, single-temperature blackbody. We overplot the model, with approximate blackbody temperatures, and normalize it to line up with the data using the normalization term $f$. Due to the uncertainties in the system parameters, we cannot reliably constrain the size of the disc from $f$. At some point during the hard state rise, the OIR colour becomes much redder than that expected from the blackbody. As soon as the source begins making the transition to the soft state, the OIR colour becomes bluer (the upper grey data in the CMD) and rejoins the expected track for the blackbody. While in the soft state, the source then follows this blackbody track as the OIR (and X-ray) flux decreases. When the soft-to-hard transition occurs the OIR counterpart again becomes redder, before decreasing in flux and rejoining the blackbody track back to quiescence. The coloured arrows in the lower left region of the CMD illustrate the tracks the data follow throughout the outburst.

The X-ray HID (Fig. 1, inset) illustrates the X-ray evolution of the source and the symbols are the same as those in the CMD. The OIR jet fades when the X-rays start to make the transition to the soft state \citep[the same has been observed many times in GX 339--4]{homaet05,coriet09}. The OIR jet reappears when the 
X-ray spectrum is hard again, once the transition back to the hard state is complete. The radio jet, with an optically thick spectrum was detected in the hard state decline once the OIR jet had recovered, and a radio detection (optically thin) was also made when the X-ray spectrum was soft but then was quenched when the source decreased in luminosity in the soft state \citep{corbet01}. For a recent discussion of the relation between radio and NIR-emitting jets in this outburst including an occasional strong jet contribution to the X-ray flux, see \cite{russet10}.

Except for the two red excursions in the CMD at the brightest times during the hard state, the blackbody model is able to approximately reproduce the observed general relationship between colour and magnitude. Emission from the jet, which is redder than the accretion disc, must be responsible for the two hard state excursions, because it is known that the jet dominated the NIR emission during these times. When the jet does not appear to dominate (during the soft state and at low flux levels in the hard state), the outer, likely X-ray heated accretion disc, which can be described as a simple blackbody of approximately constant area, dominates. We see no evidence for the spectral index to converge to $\alpha \approx +1/3$ during outburst, which would be the case in the power law spectral regime of the viscously heated disc. Nor do we see the spectral index converging to $\alpha \approx -1$, as may be expected in the power law regime of the irradiated disc \citep{franet02}. The data are unlikely to be explained by a viscously heated disc because we expect a spectral index no less than $\alpha \sim +1/3$. $\alpha < +1/3$ is observed only at low luminosities in the hard state, although due to uncertainties in the extinction we cannot rule out a viscous disc completely. However, at frequencies dominated by the Rayleigh-Jeans tail of the viscous disc blackbody, the irradiated disc contribution has been shown to be much brighter \citep[e.g.][]{hyneet02,hyne05,gieret09,gandet10}. An irradiated disc is therefore much more likely to be responsible for the observed blackbody path in the CMD than the viscous disc.

The data cannot be explained by a blackbody that changes area significantly -- this would cause the flux to change for a given temperature, which is not observed in the CMD, except possibly on the hard state rise where the data appear slightly fainter, 0.5 mag below the blackbody model. The outer disc appears to be $\sim 30$ -- 40 per cent fainter at a given temperature on the hard state rise than during the rest of the outburst. This may be due to a slightly smaller irradiated disc area \citep[e.g.][]{hyneet02} during the early stages of the outburst. Alternatively, the different normalization in the CMD could be explained by the disc being slightly bluer (hotter) at a given V mag on the hard state rise.

The results suggest that the central X-ray source is able to illuminate a $\sim$ constant area of the disc during the outburst, and that the disc is unlikely to be very warped. A warped disc would probably cause the illuminated area to change with time due to shadowing of some regions of the disc from the central X-ray source, unless the illuminating X-ray source lies at a large scale height above the disc \citep*[e.g.][]{ogildu01,foulet10}. In quiescence, the star largely dominates the OIR emission \citep{oroset02}.

It has been shown, theoretically and empirically \citep*{vanpet94,falcbi95,falcbi96,fend01,corbet03,market03,gallet03,gallet06,russet06} that the reprocessing of X-rays on the disc, and the broadband spectrum of the jet, both produce power law relations between X-ray and OIR (and radio in the case of jets) luminosities for black hole LMXBs. The former relation goes as $L_{\rm OIR,disc} \propto L_{\rm X}^{0.5}$ and the latter, as $L_{\rm OIR,jet} \propto L_{\rm Radio,jet} \propto L_{\rm X}^{0.7}$. If viable, these relations lead to an OIR jet that brightens and fades quicker than the OIR disc: $L_{\rm OIR,jet} \propto L_{\rm OIR,disc}^{1.4}$, which would explain why the jet begins to dominate the track in the CMD at higher fluxes and the disc dominates at lower fluxes.

\begin{figure*}
\centering
\includegraphics[width=12cm,angle=270]{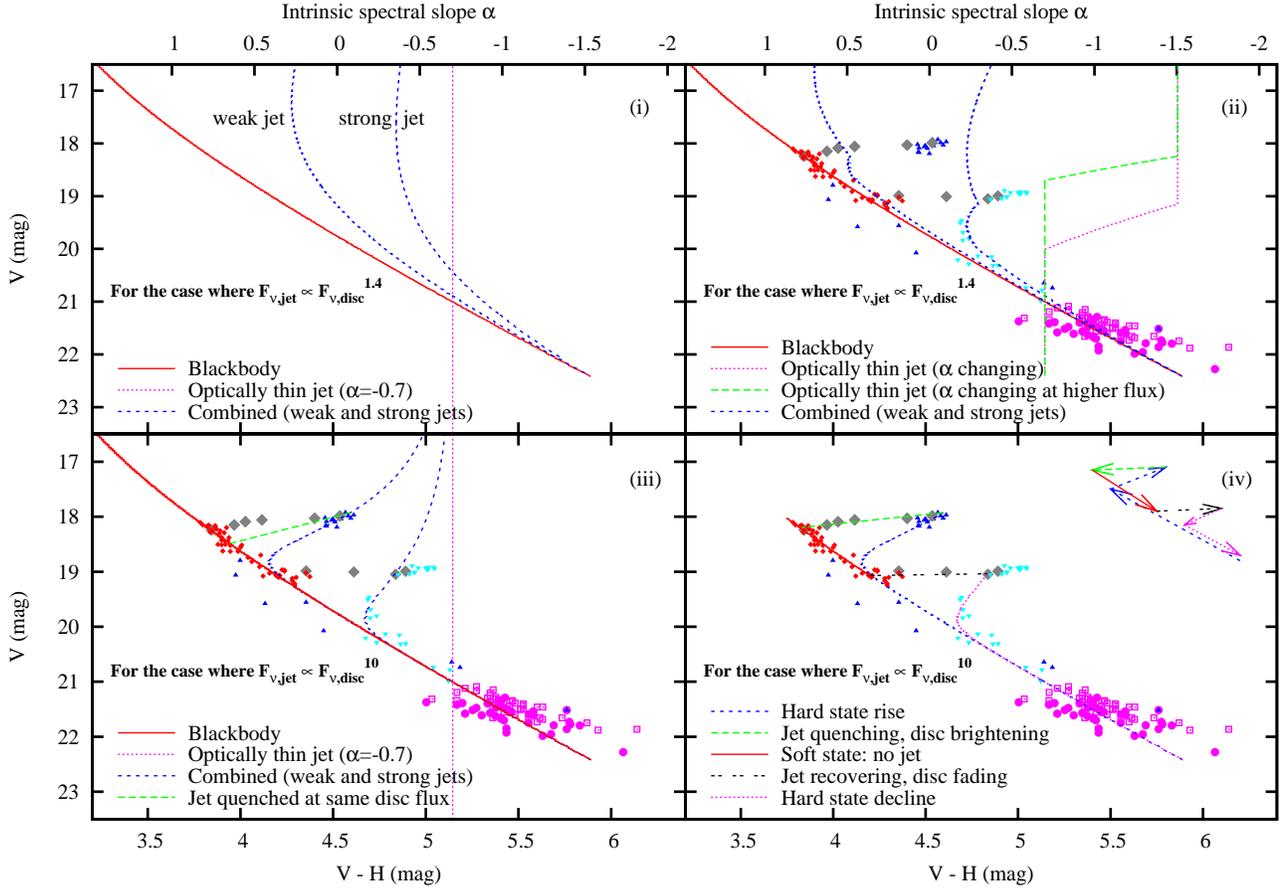}
\caption{Schematic CMDs of the XTE J1550--564 outburst. \textit{(i)} The expected track for a heated blackbody only (red solid line), an optically thin jet only (that brightens and fades slightly faster than the disc, and has a constant spectral index; magenta dotted line) and two combinations with different jet normalizations (blue dashed lines). \textit{(ii)} The same but for a changing jet spectral index, which is observed \citep{russet10}; the data are overplotted (the dotted magenta and long-dashed green lines show jets with $\alpha$ changing at different fluxes). \textit{(iii)} The same as panel (i) but for a jet which brightens and fades much more rapidly than the disc; the data are overplotted. \textit{(iv)} The same as panel (iii) but with the disc brightening/fading in transition out of/into the hard state, describing the full outburst. See the text for explanations.}
\end{figure*}

\section{Discussion of the model}

In Fig. 2 we attempt to reproduce the tracks in the CMD of XTE J1550--564 using the blackbody model and an optically thin synchrotron jet. In panel (i), we plot the expected tracks in the CMD assuming the above relations. The normalization of the $L_{\rm OIR,jet} \propto L_{\rm OIR,disc}^{1.4}$ relation could be constrained by the data. If the jet component remains a few orders of magnitude fainter than the irradiated disc component, we may expect the CMD to follow the blackbody model (red solid line); similarly if the disc is several orders of magnitude fainter than the jet, the jet emission will dominate, which we assume to be optically thin, with $\alpha = -0.7$ (magenta dotted line). The blue dashed lines in panel (i) represent intermediate cases in which the jet begins to dominate the OIR emission at the different fluxes.

These models are able to reproduce the deviation from the blackbody relation in the CMD at high flux levels, but the relative contribution of the jet compared to the disc is different on the rise and decline, because the rise jet and decline jet do not follow the same track in the CMD. For a given V magnitude, the jet normalization is higher on the decline than on the rise. This implies that at a given outer accretion disc temperature, the jet is brighter on the decline, or similarly, at a given jet flux, the temperature of the outer disc is higher on the rise. This empirical hysteresis was seen in the NIR--X-ray correlation from the same outburst \citep{russet07b}. The combined result is that the NIR jet is brighter on the decline (a) at a given X-ray luminosity, and (b) at a given outer disc temperature. The jet is either more powerful or more radiatively efficient on the outburst decline, or the viscosity parameter of the disc is changed -- any of those three possibilities could account for the hysteresis \citep{russet07b}. One clue \citep{russet10} is that the peak flux of the NIR jet is very similar on the outburst rise and decline, implying that the jet recovers to the same flux level as before the state transitions, whereas the disc and X-ray flux have faded during the soft state. This would suggest that the hysteresis may originate in the changing disc component. It is interesting to note that the NIR jet does not return to the same pre-soft state flux level in the hard state decline in GX 339--4 during three outbursts \citep{coriet09}. A NIR--X-ray hysteresis effect was also not seen in this source. It appears that, as the source fades in the soft state and returns to the hard state, the jet of XTE J1550--564 somehow retains its previous pre-soft state level of flux, but for GX 339--4 the jet does not.

\begin{figure}
\centering
\includegraphics[width=8.5cm,angle=0]{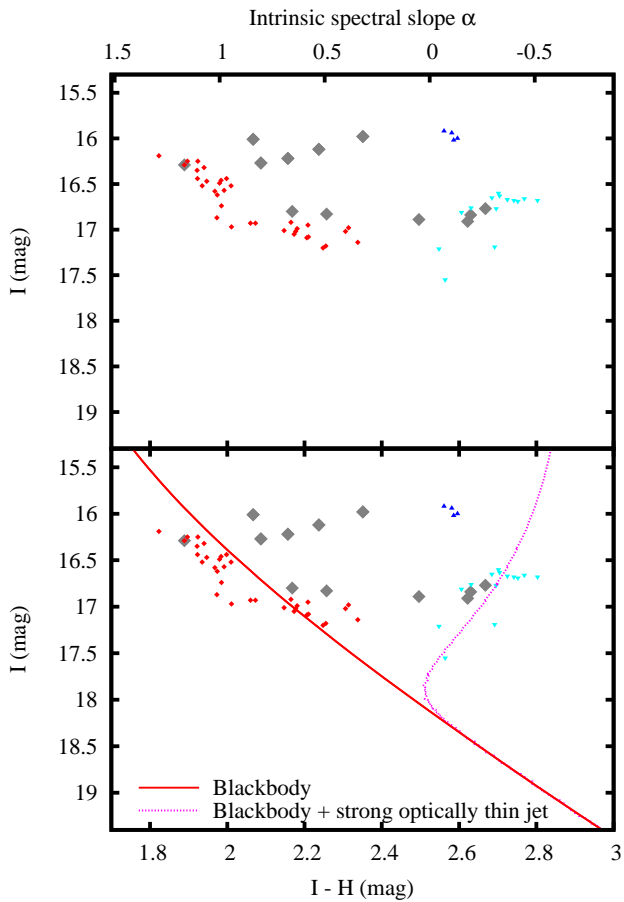}
\caption{The I--H CMD. \emph{Top panel:} The data. \emph{Lower panel:} The same model as used in Fig. 2 (\textit{iv}), showing the blackbody model and an optically thin jet, where $L_{\rm OIR,jet} \propto L_{\rm OIR,disc}^{\sim 10}$.}
\end{figure}

\subsection{A jet decoupled from the disc?}

The above models do not however reproduce the shape of the turn-off from the blackbody track in the CMD. The models predict a more gradual colour change than the data depict. The `hairpin' shape in both the rise and decline, in which the V--H colour reddens by $\sim 0.5$ mag while the magnitude brightens by $\sim 1$ mag, cannot be reproduced by the above model. In Fig. 2(ii), we show that the hairpin shape can be approximately reproduced on the outburst decline if the spectral index of the jet changes between $\alpha \sim -1.5$ and $\alpha \sim -0.6$, as is observed \citep[see Fig. 2 of][]{russet10}. We are however still unable to reproduce the hairpin on the outburst rise. This is because a weaker jet (i.e. one with a lower normalization) is required for the deviation from the blackbody track to occur at a higher flux, but with this weaker jet only a small deviation can be made. In Fig. 2(iii), we consider the possibility of a jet that brightens and fades much more rapidly than the irradiated disc. The hairpins are then reproduced if $L_{\rm OIR,jet} \propto L_{\rm OIR,disc}^{\sim 10}$. This relation is remarkably steeper than that expected. The jet flux is changing much more rapidly than the disc, in fact the results are consistent with the two being decoupled -- a large change in jet flux occurring while the irradiated disc flux is almost unchanged. In \cite{russet10} it was shown that the OIR disc component is correlated with time during the outburst fade (exponential decay) whereas the OIR jet is correlated linearly with the soft X-ray flux, and was found to fade faster than the OIR disc component.

The path XTE J1550--564 traces through its CMD is approximately reproduced by the models in Fig. 2 (\textit{iv}). $L_{\rm OIR,jet} \propto L_{\rm OIR,disc}^{\sim 10}$ is used, which reproduces the almost horizontal lines in the CMD during state transition. However, these models assume a constant spectral index for the jet, $\alpha = -0.7$; empirically we see an index which evolves with time, at least on the outburst decline \citep{russet10}. A changing spectral index is able to reproduce the hairpin in the CMD on the decline but not on the rise (Fig 2 (\textit{ii})).

We conclude that the above models can successfully reproduce the observed track in the CMD of XTE J1550--564 as long as the jet is brighter (more powerful or more radiatively efficient) on the outburst decline, but that the reddening of the colour due to the jet is much more abrupt than we would have supposed. This is likely due to a combination of an evolving optically thin synchrotron spectral index of the jet ($\alpha \sim -1.5$ to $\alpha \sim -0.7$), and a jet that brightens and fades much more rapidly than expected compared to the irradiated disc; the jet and irradiated disc may indeed sometimes be decoupled.

The disc relation \citep{vanpet94} exists globally for a collection of black hole and neutron star LMXBs \citep*{russet06,russet07a} but is not consistent with all sources individually; in another work, \cite*{dubuet01} predict a $L_{\rm OIR,disc}$--$L_{\rm X}$ relation that changes during an outburst decay from $L_{\rm OIR,disc} \propto L_{\rm X}^{\sim 0.5}$ near the outburst peak to $L_{\rm OIR,disc} \propto L_{\rm X}^{\sim 1.0}$ nearer quiescence. Theoretically, the expected jet relation of $L_{\rm OIR,jet} \propto L_{\rm Radio,jet} \propto L_{\rm X}^{0.7}$ holds if the position of the jet break between optically thick and optically thin emission lies close to the OIR regime, for all luminosities. This may be the case, but has not yet been proven \citep[see][for some recent constraints]{nowaet05,coriet09}. It is worth noting that \cite{coriet11} found $L_{\rm Radio,jet} \propto L_{\rm X}^{1.4}$ at high luminosities in the hard state of H1743--322, consistent with a radiatively efficient hard state regime. If this were the case for XTE J1550--564, we may expect $L_{\rm OIR,jet} \propto L_{\rm X}^{1.4}$, producing a more abrupt hairpin in the CMD, but still unable to account for the $L_{\rm OIR,jet} \propto L_{\rm OIR,disc}^{\sim 10}$ that can approximate the hairpin. A jet decoupled from the disc may also be consistent with observations of some BHXBs on short timescales (seconds), where the optical jet synchrotron is anti-correlated with the X-ray flux \citep[e.g.][]{kanbet01,malzet04}.

\subsection{The I--H colour-magnitude diagram}

In \cite{jainet01}, the central parts of the outburst were also monitored in I-band, which lies between V and H in wavelength. The expected wavelength dependency of the model can therefore be tested using a CMD which includes the I-band data. In Fig. 3 the CMD of the outburst is shown using I and H-band data. The intrinsic spectral index \citep{maitba08} for this CMD is $\alpha _{IH;\rm int} = 4.644 - 1.840 \times (I-H)_{\rm obs}$. Visible in the CMD is the bluer colour in the soft state, the reddening with decreasing flux in the soft state, the transitions, and the redder colour in the hard state.

The blackbody model (red solid line) lies close to the data in the soft state. The data only span $\sim 1.5$ mag in I-band (whereas the data in Fig. 1 span $\sim 4$ mag in V-band), and there is considerable scatter in the soft state I--H data around the model. Specifically, the slope in the soft state is slightly shallower than the model predicts. On the outburst decline, the optically thin jet, again probably decoupled from the disc ($L_{\rm OIR,jet} \propto L_{\rm OIR,disc}^{\sim 10}$), approximately reproduces the path the data follow, again with some scatter. The transitions are almost horizontal in the figure, as they were in Fig. 1, consistent with a fading/recovering jet while the disc brightens/fades slightly. At the peak of the hard state (both on the outburst rise and decay), the I--H spectral index is generally similar to the V--H spectral index, consistent with a power law. In the soft state the I--H spectral index appears a little bluer than the V--H spectral index, which implies some curvature to the spectrum. This is expected for a blackbody, where $\alpha$ converges to $+2.0$, the Rayleigh-Jeans tail, in the redder bands.

\section{Summary}

We have presented a simple method to successfully isolate optical/infrared disc and jet emission in an evolving X-ray transient, XTE J1550--564 using the OIR data from its 2000 outburst. Monitoring of the outburst in two wavebands is required, which emphasizes the importance of such campaigns \citep[see e.g.][]{jainet01,lewiet08,lewiet10,coriet09}. The colour-magnitude diagrams show:
\begin{itemize}
\item X-ray state dependence and hysteresis analogous to the X-ray hardness--intensity diagram;
\item a disc which can be successfully described by a heated, likely unwarped, single temperature blackbody of $\sim$ constant area, probably irradiated by X-rays (with possible evidence for a slightly smaller disc area on the outburst rise);
\item a synchrotron-emitting jet that contributes mostly at higher luminosities during the hard state and is apparently decoupled from the outer irradiated accretion disc at the high luminosities but linearly correlated with the X-ray flux at low luminosities.
\end{itemize}

\section*{Acknowledgments}

We thank Andrew King for insightful discussions regarding irradiated discs. DMR acknowledges support from a Netherlands Organization for Scientific Research (NWO) Veni Fellowship. RJHD acknowledges support from the Alexander von Humboldt Foundation.

\bsp

\label{lastpage}

\end{document}